\documentclass[12pt]{article}
\usepackage{amsmath}
\usepackage{graphicx}
\usepackage{enumerate}
\graphicspath{{figures/}}
\usepackage{natbib}
\usepackage{url}

\usepackage{bbm}
\usepackage{graphicx,psfrag,epsf}
\usepackage{enumerate}
\usepackage{natbib}
\usepackage{amsthm}
\usepackage{amssymb}
\usepackage{bm}
\usepackage{caption}
\usepackage{multirow}
\usepackage{booktabs}
\usepackage{microtype}
\usepackage{booktabs}

\usepackage{amsmath}
\usepackage{amsfonts}
\usepackage{url} 

\usepackage{mathtools}

\usepackage{color}
\usepackage[colorlinks,citecolor=blue,urlcolor=blue]{hyperref}

\usepackage{amsmath}
\usepackage{amssymb}
\usepackage{mathtools}
\usepackage{amsthm}
\usepackage{xcolor}

\usepackage{xr}
\makeatletter
\newcommand*{\addFileDependency}[1]{
	\typeout{(#1)}
	\@addtofilelist{#1}
	\IfFileExists{#1}{}{\typeout{No file #1.}}
}
\makeatother

\newcommand*{\myexternaldocument}[1]{%
	\externaldocument{#1}%
	\addFileDependency{#1.tex}%
	\addFileDependency{#1.aux}%
}
\myexternaldocument{supp-20250318}

	\newtheorem{theorem}{Theorem}
	
	\newtheorem{proposition}{Proposition}
	\newtheorem{definition}{Definition}
	\newtheorem{corollary}{Corollary}
	
	\newtheorem{remark}{Remark}
 \newtheorem{condition}{Condition}
 \newtheorem{example}{Example}

\newcommand{\anon}{1}

\begin{document}
\def\spacingset#1{\renewcommand{\baselinestretch}%
{#1}\small\normalsize} \spacingset{1}
\def\var{{\rm Var}\,}
	\def\blue#1{\textcolor{blue}{#1}}
	\def\red#1{\textcolor{red}{#1}}

\if1\anon
{
  \title{\bf Non-Asymptotic Analysis of Online Local Private Learning with SGD}
 \author{
Enze Shi\textsuperscript{1},
Jinhan Xie\textsuperscript{2},
Bei Jiang\textsuperscript{1},
Linglong Kong\textsuperscript{1},
Xuming He\textsuperscript{3}
}
\date{}
  \maketitle
} \fi

\vspace{-1cm}
\begin{center}
\textsuperscript{1}Department of Mathematical and Statistical Sciences, University of Alberta\\
\textsuperscript{2}Yunnan Key Laboratory of Statistical Modeling and Data Analysis, Yunnan University\\
\textsuperscript{3}Department of Statistics and Data Science, Washington University in St. Louis\\

\end{center}



\begin{abstract}
Differentially Private Stochastic Gradient Descent (DP-SGD) has been widely used for solving optimization problems with privacy guarantees in machine learning and statistics. Despite this, a systematic non-asymptotic convergence analysis for DP-SGD, particularly in the context of online problems and local differential privacy (LDP) models, remains largely elusive. Existing non-asymptotic analyses have focused on non-private optimization methods, and hence are not applicable to privacy-preserving optimization problems. This work initiates the analysis to bridge this gap and opens the door to non-asymptotic convergence analysis of private optimization problems. A general framework is investigated for the online LDP model in stochastic optimization problems. We assume that sensitive information from individuals is collected sequentially and aim to estimate, in real-time, a static parameter that pertains to the population of interest. Most importantly,
we conduct a comprehensive non-asymptotic convergence analysis of the proposed estimators in finite-sample situations, which gives their users practical guidelines regarding the effect of various hyperparameters, such as step size, parameter dimensions, and privacy budgets, on convergence rates.
Our proposed estimators are validated in the theoretical and practical realms by rigorous mathematical derivations and carefully constructed numerical experiments.
\end{abstract}

\noindent%
\vfill

\newpage
\spacingset{1.9}

\section{Introduction}
\renewcommand{\theequation}{1.\arabic{equation}}
\setcounter{equation}{0}

Machine learning (ML) models have proven to be highly effective in analyzing sensitive user data across various domains, including medical imaging \citep{tang2019role, kaissis2021end}, healthcare \citep{esteva2017dermatologist, wiens2019no}, finance \citep{caruana2015intelligible}, and social media content \citep{kosinski2013private, alwafi2024machine}. However, ML algorithms require extensive training data, often containing sensitive personal information, which poses significant risks to individual privacy \citep{shokri2017membership}. Differential Privacy (DP) \citep{dwork2006calibrating}, a widely accepted method for preserving privacy, provides
provable protection against identification risks. It is particularly robust to arbitrary auxiliary information that may be accessible to potential attackers. Specifically, when DP is implemented, an attacker with auxiliary data cannot infer substantially more about an individual in the database than they could if the individual's data were completely excluded from the dataset \citep{wood2018differential}.



To protect user privacy during the model training process, the widely-used approach is noisy stochastic gradient descent, commonly referred to as DP-SGD \citep{abadi2016deep}, which clips the gradient and adds carefully calibrated noise at each step of updating model parameters. This includes works that have developed privacy-preserving variants of optimization algorithms, such as performing DP-SGD with momentum \citep{kairouz2021practical, du2021dynamic}, dynamic Langevin diffusion \citep{chourasia2021differential, ryffel2022differential, ganesh2023universality}, and adaptive gradient clipping \citep{andrew2021differentially, fang2022improved}, among many others. {All these methods are designed to achieve a faster convergence rate for the central DP mechanism, assigning uniform privacy parameters to all data samples.} One shortcoming of the standard definition of DP-based algorithms is the assumption of a trusted curator who has access to the full dataset of individuals.


One way to address this issue is by allowing individuals to process their data through a DP mechanism at the local level before sending it for further processing, ensuring that the curator only receives data that have already been privatized. This approach is known as Local Differential Privacy (LDP), which has emerged as a more rigorous differential privacy technique to preserve the privacy of personal information \citep{kasiviswanathan2011can, duchi2013local}. Unlike central DP settings, where a trusted curator collects and processes the data, LDP does not rely on a trusted data curator. In this framework, each individual applies a differential privacy mechanism to their own data locally before transmitting the perturbed version to the central server. In the DP literature, this setting is referred to as the local model. The server collects the perturbed (private) data from each individual and combines them for data analysis. Due to its superior privacy protection, LDP has rapidly attracted substantial interest and found widespread applications in the industrial sector. Prominent technology companies like Google \citep{erlingsson2014rappor}, Apple \citep{tang2017privacy}, and Microsoft \citep{ding2017collecting} have already integrated LDP into their respective product portfolios.

To date, a wealth of studies on LDP have surfaced \citep{duchi2013local, duchi2018minimax, duchi2019lower, wang2019sparse, husain2020local, acharya2020context, butucea2023interactive, bhaila2024local}. Still, the local model has received considerably less attention than the central one. One contributing factor is the inherent limitations present in the local model. As a result, many fundamental problems have yet to be fully understood in the local model. In particular, a significant gap exists in research concerning the non-asymptotic convergence analysis of private stochastic approximation schemes under LDP models, especially in an online setting. Scenarios where individuals contribute data sequentially are commonplace, including situations such as web users registering on a web application after providing their information, patients being admitted to a hospital, or customers applying for a bank loan.

In this work, we are interested in stochastic optimization problems under an online LDP framework, where data from individuals are collected sequentially in a privacy-preserving manner. Specifically, our contributions can be summarized as follows:
\begin{enumerate}
 \item We present a general online locally private SGD algorithm, as well as its averaged version under the LDP model, referred to as LDP-SGD and LDP-ASGD. To avoid the limitation of traditional DP-SGD \citep{dwork2010boosting,dong2022gaussian}, where training for a large number of iterations results in high privacy costs \citep{avella2023differentially} due to the sequential composition property, our strategy combines a one-pass estimation procedure without re-accessing the historical data, leveraging the benefits of the parallel composition property of private algorithms.
    \item We conduct a systematic non-asymptotic convergence analysis to better understand the privacy-utility trade-off with LDP-SGD and LDP-ASGD estimators. In particular, our work focuses on estimating the parameter of interest for the population from which sensitive data are being collected, within the framework of convex stochastic optimization problems. 
        \item A further contribution of our work is to show that both LDP-SGD and LDP-ASGD estimators display the same sensitivity to parameter dimension and differential privacy budget. However, they exhibit divergent convergence rates under varying step-size configurations. We provide a comprehensive analysis of the convergence rates in different settings, and summarize the results in Section \ref{sec:summary}.
\end{enumerate}

 The structure of this paper is organized as follows: In Section \ref{sec:pre}, we present some basic concepts of DP and LDP. We introduce our proposed methodology in Section \ref{sec:meth}, delving into the LDP-SGD and LDP-ASGD algorithms. The non-asymptotic theoretical analyses are presented in Section \ref{sec:thm}. We conclude by presenting experimental results in Section \ref{sec:experiment}, which underscore the efficacy of our approach. In Section \ref{sec:real}, we apply the proposed method to the US insurance data. Some conclusions are provided in Section \ref{sec:con}. To conserve space, additional simulation studies and detailed proofs of theorems are included in the Supplementary Material.

\section{Preliminaries}
\label{sec:pre}

In this section, we briefly introduce some background knowledge about DP and LDP.

\begin{definition}[($\varepsilon,\delta$)-DP \citep{dwork2006calibrating}]
{Let $\mathcal{X}$ be the sample space for individual data}. A randomized algorithm ${M}$: $\mathcal{X}^n \rightarrow \mathcal{R}$, is $(\varepsilon,\delta)$-differentially private if and only if for every pair of adjacent datasets $\boldsymbol{X}$  $\boldsymbol{X}^{\prime}\subset\mathcal{X}^n$ and for any measurable event $E \subseteq \mathcal{R}$, the inequality below holds:
$$
{\rm{Pr}}(M(\boldsymbol{X}) \in E) \leq e^{\varepsilon} \cdot {\rm{Pr}}(M(\boldsymbol{X}^{\prime}) \in E)+\delta,
$$
where we say that two datasets $\boldsymbol{X} = \{x_i\}_{i=1}^n$ and $\boldsymbol{X}^\prime = \{x_i^\prime\}_{i=1}^n$ are adjacent if and only if they differ by one individual datum and the probability measure $\rm{Pr}$ is induced by the randomness of $M$ only. When $\delta=0$, then $M$ is called $\varepsilon$-differentially private ($\varepsilon$-DP).
\end{definition}

{From this definition, we know that the parameters $\varepsilon,\delta$ control the privacy level. As both $\varepsilon$ and $\delta$ decrease, the outputs for adjacent datasets, $\boldsymbol{X}$ and $\boldsymbol{X}^\prime$, become closer, making it more difficult for an adversary to distinguish between the two datasets. This signifies that the privacy guarantee becomes stronger as these parameters decrease.} 
While the concept of $(\varepsilon,\delta)$-DP has broad applicability in various domains such as healthcare, finance, and social networks, there are growing concerns regarding a potential vulnerability in the process: the reliance on a trusted curator. In the following, we introduce a more rigorous LDP concept for scenarios where the data collector at the central server cannot be trusted.

\begin{definition}[($\varepsilon,\delta$)-LDP~\citep{ xiong2020comprehensive}]
   A randomized algorithm $M: \mathcal{X} \rightarrow \mathcal{R}$ is $(\varepsilon,\delta)$-local differential privacy ($(\varepsilon,\delta)$-LDP), where $\varepsilon \geq 0$ and $0\leq \delta\leq 1$, if and only if any pair of input individual values $x, x^\prime \in \mathcal{X}$, for every measurable event $E \subseteq \mathcal{R}$,
   $$
{\rm{Pr}}(M(x) \in E) \leq e^{\varepsilon} \cdot {\rm{Pr}}(M(x^{\prime}) \in E)+\delta.
$$
If $\delta =0$, the algorithm $M$ satisfies pure (strict) local differential privacy (pure LDP), namely, $\varepsilon$-LDP.
\end{definition}
{LDP has two main frameworks for privacy preservation: interactive and non-interactive frameworks \citep{duchi2018minimax, butucea2023interactive}. Meanwhile, similar to the centralized model, LDP controls plausible deniability for any two individual datum by ensuring that the algorithm $M$ satisfies $\varepsilon$-LDP or $(\varepsilon,\delta)$-LDP.} In a local setting, each participant autonomously manages his/her data, independently. Correspondingly, the responsibility for privacy processing shifts from a data collector (central) to each participant (local). Subsequently, we present the concept of sensitivity, parallel composition property, and post-processing property, which enables us to design the proposed private algorithm with streaming data.
\begin{definition}[$\ell_p$-sensitivity] Let $f: \mathcal{X}^n \rightarrow \mathcal{R}^d$ be a query mapping. For a fixed positive scalar $p$, the $\ell_p$-sensitivity of $f$ is defined by  $\Delta(f) = \sup_{\boldsymbol{X},{\boldsymbol{X}}^\prime\subset\mathcal{X}^n, adjacent}\|f(\boldsymbol{X})-f({\boldsymbol{X}}^\prime)\|_p$.
\end{definition}

\begin{proposition}[Parallel composition~\citep{xiong2020comprehensive}]
\label{lemma2}
Suppose there are $n$ mechanisms $\{M_1,\ldots,M_n\}$ satisfy
$(\varepsilon_i,\delta_i)$-LDP, respectively, and are computed on a disjoint subset of the private data, then a mechanism formed by $(M_1(x_1),\cdots,M_n(x_n))$ satisfies $(\max\varepsilon_i, \max\delta_i)$-LDP.
\end{proposition}

\begin{proposition}[Post-processing property~\citep{xiong2020comprehensive}]
\label{lemma3}
If mechanism $M_1$ satisfies $(\varepsilon,\delta)$-LDP, for any mechanism $M_2$, even may not satisfies LDP, the composition of $M_1$ and $M_2$, namely, $M_2(M_1(\cdot))$ satisfies $(\varepsilon,\delta)$-LDP.
\end{proposition}

\section{Methodology}
\label{sec:meth}
\renewcommand{\theequation}{3.\arabic{equation}}
\setcounter{equation}{0}

Let $x_1,\ldots,x_n$ be independent and identically distributed observations from the probability distribution $\Pi$, representing the private information of each user. For statistical estimation and machine learning problems, we consider the general situation where the true $d$-dimensional model parameter $\theta^\ast\in\mathcal{R}^d$ is the minimizer of a convex objection function  $F(\theta)$: $\mathcal{R}^d\rightarrow\mathcal{R}$, that is,
\begin{align*}
  \theta^\ast =\operatorname{argmin}\left\{ F(\theta) = {E}_{x\in\Pi}f(\theta,x)\right\},
\end{align*}
where $x$ is a random variable drawn from a probability distribution $\Pi$ and $f(\theta,x)$ is the loss function such as the {squared, Huber and cross-entropy loss.} 
A widely used optimization method for minimizing $F(\theta)$ {with streaming data} is the stochastic gradient descent \citep{robbins1951stochastic, polyak1992acceleration}. In particular, let $\theta_0$ be any given initial point. Recall that the {online} SGD updates the iterate as follows:
\begin{align}{\label{sgd}}
    \theta_i = \theta_{i-1} - \eta_i\nabla f(\theta_{i-1},x_i),\quad i\geq 1,
\end{align}
where $x_i$ is the $i$th observation, 
$\nabla f(\theta,x)$ denotes the gradient of $f(\theta,x)$ with respect to $\theta$, and $\eta_i$ is the $i$th step size, which we refer to as the learning rate.

Within the framework of streaming data, we introduce a locally differentially private estimator utilizing noisy SGD. Intuitively, the heightened risk of privacy leakage stems from the numerous data and gradient queries inherent in the SGD algorithm. To mitigate this, we utilize LDP to achieve privacy by restricting the class of estimators to satisfy a uniform boundedness condition.

\begin{condition}{\label{Con1}}
The gradient of the loss function $f(\theta,x)$ is such that $\sup_{\theta\in\mathcal{R}^d,x\in\Pi}\|\nabla f(\theta, x)\|_2\leq C_0<\infty,$
where $C_0$ is some positive constant.
 \end{condition}

\begin{figure}[h]
  \centering
\includegraphics[width=3.5cm]{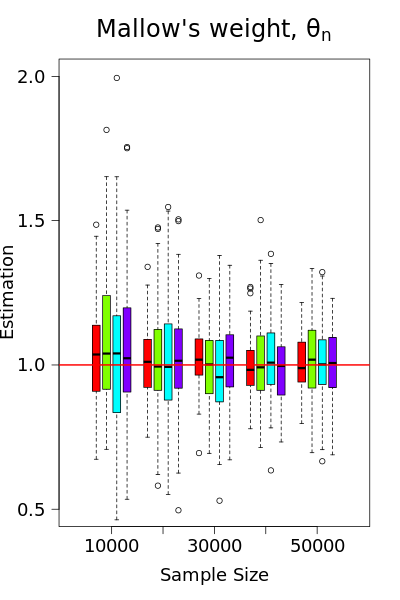}
\includegraphics[width=3.5cm]{{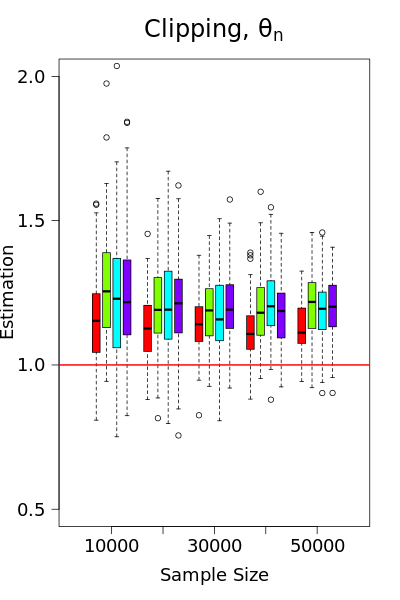}}
\includegraphics[width=3.5cm]{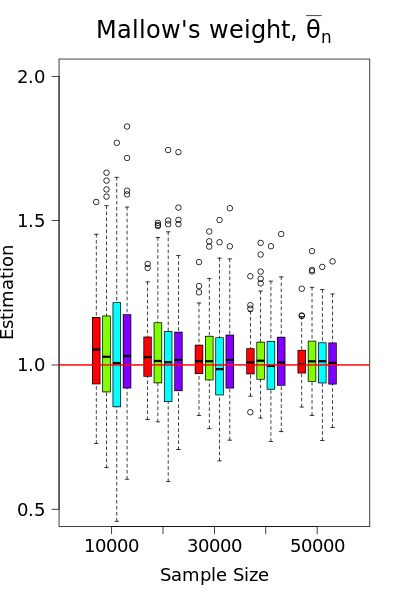}
\includegraphics[width=3.5cm]{{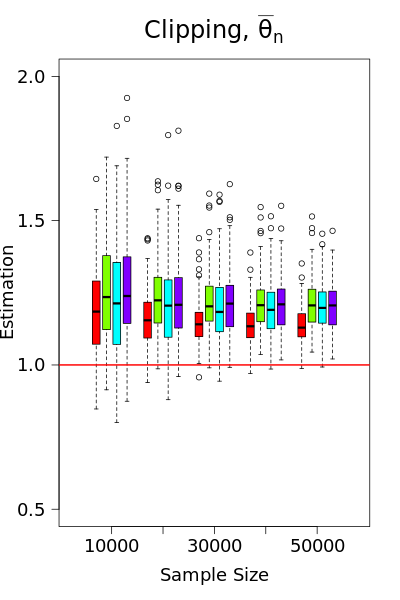}}
  \caption{Boxplots comparing LDP-SGD and LDP-ASGD estimators between Mallows weights and gradient clipping for logistic regression with dimension $d=3$. The red line indicates the true parameter $\theta^*=(1,1,1,1)$ and the boxplots represent the estimators from 100 replications. The four colors correspond to four different coordinates of the parameters. The clipping method results in a positive bias, which does not arise with the Mallows weights method. The details for data generation can be found in the Supplementary Material.}
  \label{fig:clip}
\end{figure}

\begin{remark}
    The bounded gradient assumption plays a critical role in ensuring sensitivity control during iterations within the differential privacy framework. This assumption can be readily met by applying strategies such as Mallows weights \citep{loh2017statistical} or gradient clipping \citep{abadi2016deep}, as discussed in \cite{avella2023differentially}. However, as highlighted in Figure \ref{fig:clip}, employing gradient clipping introduces inconsistencies in the estimators. This challenge underscores our preference for using Mallows weights alongside the assumption of a bounded gradient.
\end{remark}
Condition \ref{Con1} implies that the objective loss function $f(\theta,x)$ possesses a predefined gradient bound, denoted as $C_0$.  This ensures that the sensitivity of $\nabla f(\theta,x)$ does not exceed $2C_0$ when considering $\ell_2$-sensitivity. Therefore, to achieve a certain level of privacy, our locally private version of SGD considers the following noisy version of the iterates (\ref{sgd}), at any given initial point  $\theta_0$:
\begin{align}{\label{psgd}}
    {\theta}_i =\theta_{i-1} -\eta_i\nabla f({\theta}_{i-1},x_i) + \eta_i C_0\omega_i(\varepsilon_i,\delta_i),\quad i\geq 1,
\end{align}
where $\{\omega_i(\varepsilon_i,\delta_i)\}_{i\geq 1}$ represents a sequence of i.i.d. $d$-dimensional noise vectors, depending on the individual's certain level of privacy budget $(\varepsilon_i,\delta_i)$. Taking $C_0=0$ in (\ref{psgd}) recovers the standard SGD algorithm in (\ref{sgd}). We present the following three typical examples to illustrate specific forms of $\omega_i(\varepsilon_i,\delta_i).$

\begin{example}(Laplace mechanism, \cite{dwork2014algorithmic})
Under Condition \ref{Con1}, and by applying the Cauchy–Schwarz inequality, we know that
 \begin{align*}
     \|\nabla f(\theta, x) - \nabla f(\theta, x^\prime)\|_1 \leq \sqrt{d}\|\nabla f(\theta, x) - \nabla f(\theta, x^\prime)\|_2
 \end{align*}
 for two different values $x$ and $x^\prime.$ Correspondingly, the $\ell_1$-sensitivity of $\nabla f(\theta,x)$ is bounded by $2\sqrt{d}C_0.$ Therefore, $\theta_i$ achieves $\varepsilon_i$-LDP if {each component of the $d$-dimensional noise vector} $\omega_i(\varepsilon_i,\delta_i)$ is sampled from ${\rm Laplace}(0,2\sqrt{d}/\varepsilon_i)$.
\end{example}

\begin{example}(Gaussian mechanism, \cite{dwork2014algorithmic})
If instead we consider $\ell_2$-sensitivity of $\nabla f(\theta,x)$, for $d$-dimensional noise vectors $\omega_i(\varepsilon_i,\delta_i)$ sampled from normal distribution ${N}(\boldsymbol{0}_d, (8\log(1.25/\delta_i)/\varepsilon_i)\boldsymbol{I}_d)$, then $\theta_i$ achieves $(\varepsilon_i, \delta_i)$-LDP.
\end{example}

\begin{example}(Gaussian differential privacy mechanism, \cite{dong2022gaussian}) \label{eg3}
To overcome the weak interpretability of $(\varepsilon,\delta)$-DP, GDP provides an intuitively appealing interpretation: determining whether an individual is part of a given dataset is at least as challenging as distinguishing between two normal distributions--specifically ${N}(0,1)$ and ${N}(\mu,1)$ based on one draw, where $\mu \geq 0$. For $\omega_i(\varepsilon_i,\delta_i)$ sampled from ${N}(\boldsymbol{0}_d, (4/\mu_i)\boldsymbol{I}_d)$, then $\theta_i$ achieves local $\mu_i$-GDP. Actually, using Corollary 1 in \citet{dong2022gaussian}, $\theta_i$ is also $(\varepsilon_i,\delta_{\mu_i}(\varepsilon_i))$-LDP for all $\varepsilon_i>0$, where
 \begin{align*}
     \delta_{\mu_i}(\varepsilon_i) = \Phi(-\frac{\varepsilon_i}{\mu_i} + \frac{\mu_i}{2}) - e^{\varepsilon_i}\Phi(-\frac{\varepsilon_i}{\mu_i} - \frac{\mu_i}{2})
 \end{align*}
and $\Phi(\cdot)$ is the cumulative distribution function of the standard normal
distribution.
\end{example}

 From (\ref{psgd}), each individual has different privacy budgets $(\varepsilon_i,\delta_i)$, $i=1,\ldots, n$, which protect the privacy of individual data owners without the need for a trusted third party.
 In practice, stochastic approximation schemes are sequentially interactive cases, which is a typical example of LDP \citep{duchi2018minimax}. The estimator $\theta_i$ defined in (\ref{psgd}) is referred to LDP-SGD estimator. Without loss of generality, we assume each iteration of (\ref{psgd}) satisfies $(\varepsilon_i,\delta_i)$-LDP and establish the following results.

\begin{theorem}{\label{sgd11}}
  The LDP-SGD estimator  ${\theta}_i$
 obtained from (\ref{psgd}) is $(\max_{1\leq j\leq i}\varepsilon_j,\max_{1\leq j \leq i}\delta_j)$-LDP, where $i\geq 1$.
    \end{theorem}

A special case is that the proposed LDP-SGD estimator $\theta_n$ is central $(\varepsilon,\delta)$-DP when $\varepsilon_1=\cdots=\varepsilon_n = \varepsilon$ and $\delta_1 = \cdots = \delta_n = \delta$ without considering different privacy budget for each individual.

Meanwhile, we also consider the Polyak-Ruppert averaging private estimator $\bar{\theta}_n = \sum_{i=1}^n{\theta}_i/n$ after $n$ iterations, which is referred to LDP-ASGD estimator. Note that this estimator
can also be recursively updated: $\bar{\theta}_n = (n-1)\bar{\theta}_{n-1}/n + {\theta}_n/n$. The post-processing property in Proposition \ref{lemma3} along with Theorem \ref{sgd11} easily imply the following results:
\begin{theorem}{\label{asgd}}
 The LDP-ASGD estimator    $\bar{\theta}_i$ is $(\max_{1\leq j\leq i}\varepsilon_i,\max_{1\leq j \leq i}\delta_i)$-LDP.
    \end{theorem}

\begin{figure}[ht]
\centering
  \includegraphics[width=0.45\textwidth]{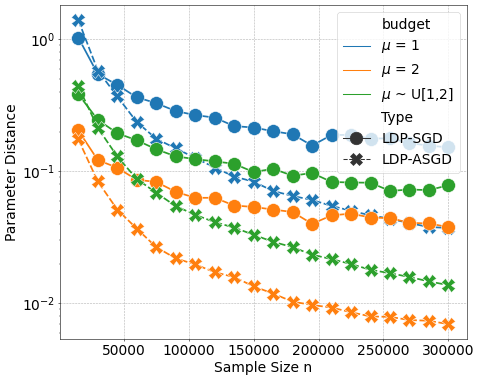}
  \includegraphics[width=0.45\textwidth]{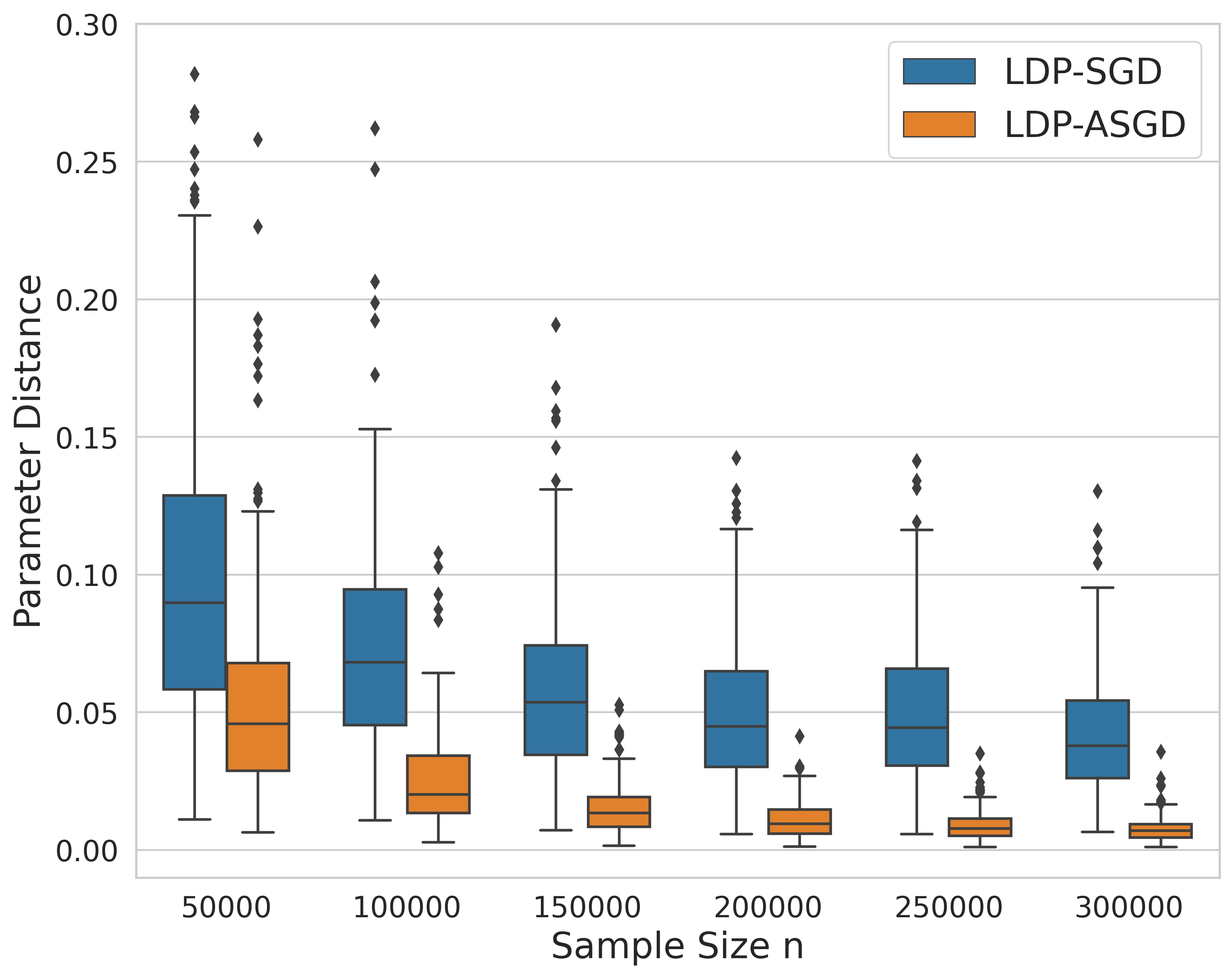}
  \caption{Trajectories of the distance between DP-SGD estimators and the ground truth for linear regression. \textbf{Left:} Trajectories for two types of estimators under varying privacy budgets with results averaged over 200 replicates. The parameter distance for privacy budgets sampled from ${\rm{Unif}}(1,2)$ lies between the cases of $\mu=1$ and $\mu=2$. \textbf{Right:} Boxplot comparison of the LDP-SGD and LDP-ASGD estimators for $\mu = 2$ over 200 replicates. The LDP-ASGD estimator demonstrates consistently lower variability and smaller parameter distance compared to LDP-SGD, especially as the sample size increases.
}
  \label{fig:intro}
\end{figure}

A visualization of the motivation for this paper is displayed in Figure \ref{fig:intro}. Referring to the left panel of Figure \ref{fig:intro}, we permit each user to assign individual privacy budgets, which aligns more closely with real-world scenarios. Meanwhile, as illustrated in the right panel of the same figure, the LDP-ASGD estimator exhibits greater stability during the iteration compared to the LDP-SGD estimator.

\section{Theoretical properties}
\label{sec:thm}
\renewcommand{\theequation}{4.\arabic{equation}}
\setcounter{equation}{0}

\subsection{Strongly convex objectives}
In this subsection, we focus on conducting a direct non-asymptotic analysis of the LDP-SGD and LDP-ASGD estimators. To better illustrate the theoretical properties of the proposed estimators, we impose the following regularity conditions on the objective functions.

\begin{condition}{\label{Con2}}
Assume that $F(\theta)$ is differentiable, $M$-smooth, and $m$-strongle convex, i.e.
\begin{itemize}
 \item[(i).]  $F(\theta_1) - F(\theta_2) \leq \langle \nabla F(\theta_2), \theta_1-\theta_2 \rangle + \frac{M}{2}\|\theta_1-\theta_2\|^2,\quad \forall\theta_1, \theta_2\in{\Theta}\subseteq\mathbb{R}^d,$
  \item[(ii).]    $F(\theta_1) - F(\theta_2) \geq \langle \nabla F(\theta_2), \theta_1-\theta_2 \rangle + \frac{m}{2}\|\theta_1-\theta_2\|^2,\quad \forall\theta_1, \theta_2\in{\Theta}\subseteq\mathbb{R}^d,$
\end{itemize}
for some constants $M$ and $m$.
 \end{condition}
Strong convexity and smoothness are both standard conditions for the convergence analysis of stochastic gradient optimization methods; similar conditions can be found in \cite{gower2019sgd, vaswani2022towards}. To establish our results, we also require the local linearity condition on the gradient of the objective function $f(\theta,\cdot)$, which is commonly assumed for SGD’s method with the Polyak-Ruppert averaging, see \cite{polyak1992acceleration,fang2018online,wang2021convergence}.
\begin{condition}{\label{Con3}}
Assume that the gradient $E\{\nabla f(\theta,\cdot)\}$ can be approximated using the Hessian matrix $E\{\nabla^2 f(\theta^\ast,\cdot)\}$  around the minimizer $\theta^\ast$ as
    \begin{align*}
    \|E\{\nabla f(\theta,x_i)\} - E\{\nabla^2 f(\theta^\ast,x_i)\}(\theta -\theta^\ast)\| \leq C_1\|\theta - \theta^\ast\|^2.
\end{align*}
 \end{condition}

Before presenting the non-asymptotic bounds, we first introduce the following family of functions: $\psi_\beta: (0,+\infty)\rightarrow \mathbb{R}$ given by:
$$
\psi_\beta(t)= (t^\beta-1)/\beta,\quad  \text { if } \beta \neq 0,~{t>0.}
$$
Notice that the function $\beta \mapsto \psi_\beta(t)$ is continuous for all $t>0$. In addition, for $\beta>0, \psi_\beta(t)<t^\beta/\beta$, while for $\beta<0$, we have $\psi_\beta(t)< -1/\beta$.

Denote $\Omega$ as the maximum expectation of the  $\ell_2$-norm of the noise vector $\omega_i(\varepsilon_i,\delta_i)$, i.e., $\Omega = \max_i\{E(\|\omega_i(\varepsilon_i,\delta_i)\|_2^2)\}$. We then have the following non-asymptotic bounds for the LDP-SGD and LDP-ASGD estimators. For notation simplicity, let $\delta_n = E(\|{\theta}_n - \theta^\ast\|^2)$, $\bar{\delta}_n = E(\|{\bar{\theta}}_n - \theta^\ast\|^2)$, $\Delta_n=E\{\hat{F}(\theta_n) - \hat{F}({\theta}^\ast)\}$ and $\bar{\Delta}_n=E\{\hat{F}(\bar{\theta}_n) - \hat{F}({\theta}^\ast)\}$, where $\hat{F}(\theta) = \sum_{i=1}^n f(x_i,\theta)/n$.

Also, for two sequences $\{a_n\}$ and $\{b_n\}$, we denote $a_n = o(b_n)$ if $a_n$ is of smaller order than $b_n$, i.e., $a_n / b_n \to 0$ as $n \to \infty$. Similarly, we denote $a_n = \mathcal{O}(b_n)$ if $a_n$ is of the same order as $b_n$, meaning that $|a_n / b_n|$ is bounded for sufficiently large $n$. For stochastic sequences, we use $\mathcal{O}_p$ and $o_p$ to denote orders in probability. Specifically, $a_n = o_p(b_n)$ if $a_n / b_n \xrightarrow{p} 0$, and $a_n = \mathcal{O}_p(b_n)$ if $a_n / b_n$ is stochastically bounded, i.e., $\sup_n P(|a_n / b_n| > M) \to 0$ as $M \to \infty$.

\begin{theorem}{\label{the1}}
Assume Conditions \ref{Con1}, \ref{Con2}, \ref{Con3} hold, if $\eta_i=\eta i^{-\alpha}$, we then have, for $\alpha\in[0,1)$
\begin{align*}
\delta_n \leq C_0^2(1+\Omega)\frac{\eta}{m n^{\alpha}} + r_n,
\end{align*}
where $r_n=o(n^{-\alpha})$ is the reminder term. The exact upper bound can be found in the proofs of the Appendix.
\end{theorem}

The quantity \( \Omega \) is a constant with respect to the sample size \( n \), but it depends on the dimension \( d \) and the privacy budget. In this theorem and those that follow, \( \Omega \) varies depending on the choice of noise distribution. Specifically, for the three mechanisms discussed earlier (Laplace, Gaussian, and GDP mechanisms), the corresponding forms of \( \Omega \) are \( \mathcal{O}(d^3 (\ln d)^2 / \min_i\{\varepsilon_i^2\}) \), \( \mathcal{O}(d / \min_i\{\varepsilon_i^2\}) \), and \( \mathcal{O}(d / \min_i\{\mu_i^2\}) \), respectively, where \( \mathcal{O}(\cdot) \) hides a positive constant independent of \( d \).

\begin{remark}\label{rmk1}
From Theorem \ref{the1}, we know that the non-asymptotic upper bound of the LDP-SGD estimator ${\theta}_n$ is influenced by several factors: the initial condition \(\|{\theta}_0 - \theta^\ast\|^2\), which is included in the reminder term $r_n$, the initial step size \(\eta\), the decay rate \(\alpha\), and the dimensionality \(d\) of the parameters. Specifically, \\[-1.5em]
\begin{itemize}
    \item \textbf{Initial condition:} The term involving the initial condition diminishes at a sub-exponential rate as \( n \) increases for \( \alpha \in [0, 1) \), which follows directly from the exact upper bound provided in the Appendix.\\[-1.5em]
    \item \textbf{Convergence rate:} The predominant convergence rate is \(\mathcal{O}(n^{-\alpha})\) for \(\alpha \in [0,1)\). For a scenario where \(\alpha=1\), the convergence of the LDP-SGD estimator \(\theta_n\) is not assured.\\[-1.5em]
    \item \textbf{Parameter dimension:} The influence of the dimensionality $d$ is implicitly captured through the maximum expectation $\Omega$. As outlined in Theorem \ref{the1}, under Condition \ref{Con1}, the Laplace noise exhibits a higher order dependence on $d$ compared to the Gaussian noise, consequently, it is more significantly affected by the curse of dimensionality. \\[-1.5em]

\end{itemize}
\end{remark}

The following theorem presents the non-asymptotic upper bounds for the LDP-ASGD estimator.

\begin{theorem}{\label{the2}}
Assume Conditions \ref{Con1}, \ref{Con2}, \ref{Con3} hold, if $\eta_i=\eta i^{-\alpha}$, we then have, for $\alpha\in[0,1)$
\begin{align*}
\bar{\delta}_n & \leq \frac{1}{n}\operatorname{tr}\left[E\{\nabla^2 f\left(\theta^*, x_1\right)\}^{-1}S E\{\nabla^2 f\left(\theta^*, x_1\right)\}^{-1}\right] \\
& \quad+ \frac{C_0^2(1+\Omega)}{mn}+ \frac{C_1 C_{3,0}\eta^2}{nm^2}\psi_{1-2\alpha}(n) +r_n, 
\end{align*}
where $x_1$ is the first sample, $S$ is a constant matrix defined latter in \eqref{eq:clt}, $C_1$, $C_{3,0}$ are the constants defined in the supplementary, and $r_n=o((1+\psi_{1-\alpha}(n))/n)$ is the reminder term. The exact upper bound can be found in the proofs of the Appendix.
\end{theorem}

\begin{remark}
   From Theorem \ref{the2}, notice that the non-asymptotic upper bound of the LDP-ASGD estimator $\bar{\theta}_n$ is influenced by the same factors in a similar way to the bound in Theorem \ref{the1}. \\[-1.5em]
    \begin{itemize}
        \item \textbf{Initial Condition:} For all $\alpha\in[0,1)$, the term involving the initial condition diminishes at rate $O(n^{-2})$, which can be obtained from the exact upper bound.\\[-1.5em]
        \item \textbf{Convergence rate:} For $\alpha\in[0,1/2]$, we have $\psi_{1-2\alpha}(n)=\mathcal{O}(n^{1-2\alpha})$, then the third term on the error bound in Theorem \ref{the2} is the asymptotic leading term with order $\mathcal{O}(n^{-2\alpha})$. For $\alpha\in[1/2,1)$, we have $\psi_{1-2\alpha}(n)=\mathcal{O}(1)$ and the third term has the same convergence rate $\mathcal{O}(n^{-1})$ as the first two terms.
    \end{itemize}
\end{remark}

\begin{corollary}(Asymptotic properties) \label{coro1}
Based on the decomposition shown in the Appendix, $\bar{\theta}_n-\theta^*$ can be represented as follows:
$$
\sqrt{n}(\bar{\theta}_n-\theta^*) = -\frac{1}{\sqrt{n}}\sum_{j=1}^n A^{-1}\left(\nabla f\left(\theta^*, x_j\right)+C_0\omega_j(\varepsilon_j,\delta_j)\right)+\mathcal{O}_p\left(\frac{\psi_{1-\alpha}(n)}{\sqrt{n}}\right)+o_p(1),
$$
where $A=E\{\nabla^2f\left(\theta^*, x_n\right)\}$. For $1/2<\alpha<1$, we have $\mathcal{O}_p(\psi_{1-\alpha}(n)/\sqrt{n})=\mathcal{O}_p(n^{1/2-\alpha})=o_p(1)$. Then by the central limit theorem, we have
\begin{align}
    \begin{aligned}
    \label{eq:clt}
        \sqrt{n}(\bar\theta_n - \theta^*)\rightarrow \mathcal{N}(0, A^{-1}SA^{-1}),
    \end{aligned}
\end{align}
where $S=\mathbb{E}\{\nabla f(\theta^*,x_n)\nabla f(\theta^*,x_n)^\top\}+S_{noise}$, and $S_{noise} = C_0^2\mathbb{E}\{\omega_n(\varepsilon_n,\delta_n)\omega_n(\varepsilon_n,\delta_n)^\top\}$. The first term in $S$ is what we can obtain from a non-private SGD algorithm, while $S_{noise}$ is the price we pay to achieve the desired level of privacy.

\end{corollary}

\subsection{Non-strongly convex obejectives}
In this subsection, we do not assume that the function $F(\theta)$ is strongly convex, but impose the following condition.
\begin{condition}{\label{Con4}}
 Assume the objective function $F(\theta)$ attains its global minimum at $\theta^\ast\in\Theta.$
\end{condition}
Notice that $\theta^\ast$ is not necessarily unique when the objective function $F(\theta)$ is not strongly convex, but we may bound the loss function as follows:

\begin{theorem}{\label{the3}}
Assume Conditions \ref{Con1}, \ref{Con2}, \ref{Con3}, \ref{Con4} hold, then if $\eta_i = \eta i^{-\alpha}$, for $\alpha\in[1/3,1)$,
\begin{align*}
\Delta_n\leq \left\{\begin{array}{lr}
{2G^{1/2}H}/{\psi_{1-\alpha}(n)},& \alpha\in(2/3,1),\\
{2G^{1/2}H}/{\psi_{\alpha/2}(n)},&\alpha\in(1/2,2/3],\\
{2JH}/\{(1-2\alpha)^{1/2}\psi_{3\alpha/2-1/2}(n)\},& \alpha\in[1/3,1/2],
\end{array}
\right.
\end{align*}
where $G = \delta_0 + C_0^2(1+\Omega)\eta^2\psi_{1-2\alpha}(n)$, $H = (1+4M^{1/2}C_0(1+\Omega)^{1/2}\eta^{3/2})/\eta$, and $J = \delta_0 +C_0^2(1+\Omega)\eta^2$.
\end{theorem}

\begin{remark}
   From Theorem \ref{the3}, we note that the upper bound of $\Delta_n$ shares the same constant $\Omega$ as in the strongly convex scenario, suggesting a similar relationship with dimension and privacy budget as discussed before. Moreover, we have \\[-1.5em]
    \begin{itemize}
        \item \textbf{Initial Condition:} For \( \alpha \in (1/3, 1) \), the initial condition \( \delta_0 \) appears in the expressions for \( G \) and \( J \). The terms involving \( \delta_0 \) decay to zero as \( n \) increases, and they dominate the convergence behavior and constitute the leading term in this regime.\\[-1.5em]
        \item \textbf{Convergence rate:} For $\alpha\in(1/3,1/2]$, the convergence rate of $\Delta_n$ is $O(n^{-(3\alpha-1)/2})$;  for $\alpha\in(1/2,2/3]$, the convergence rate is $O(n^{-\alpha/2})$, while for $\alpha\in(2/3,1)$ the convergence rate becomes $O(n^{-(1-\alpha)})$.
    \end{itemize}
\end{remark}

The following theorem is shown in a similar way to Theorem \ref{the2}, which provides the bound of the difference between the expected loss of LDP-ASGD estimators and optimal.

\begin{theorem}{\label{the4}}
Assume Conditions \ref{Con1}, \ref{Con2}, \ref{Con3}, \ref{Con4} hold, then if $\eta_n = \eta n^{-\alpha}$, for $\alpha\in[0,1)$, we have
\begin{align*}
\bar{\Delta}_n \leq \frac{1}{2\eta n^{1-\alpha}}\{\delta_0 + C_0^2(1+\Omega)\eta^2\psi_{1-2\alpha}(n)\} + \frac{C_0^2(1+\Omega)\eta\psi_{1-\alpha}(n)}{2n}.
\end{align*}
\end{theorem}
\begin{remark}
    The convergence rate of $\bar{\Delta}_n$ follows \(O(n^{-\alpha})\) when \(\alpha \in [0, 1/2)\). However, in the range \(1/2 < \alpha < 1\), it adheres to \(O(n^{\alpha-1})\). Consequently, the optimal asymptotic rate stands at \(O(n^{-1/2})\) when taking $\alpha=1/2$. It is worth noting that through averaging, the algorithm extends the \(\alpha\) range to ensure convergence, shifting it from \([1/3, 1)\) to \([0,1)\), and also achieves faster convergence rate than $\Delta_n$. 
\end{remark}

\subsection{Summary of theoretical results}
\label{sec:summary}

\begin{figure}
\centering
  \includegraphics[width=0.6\textwidth]{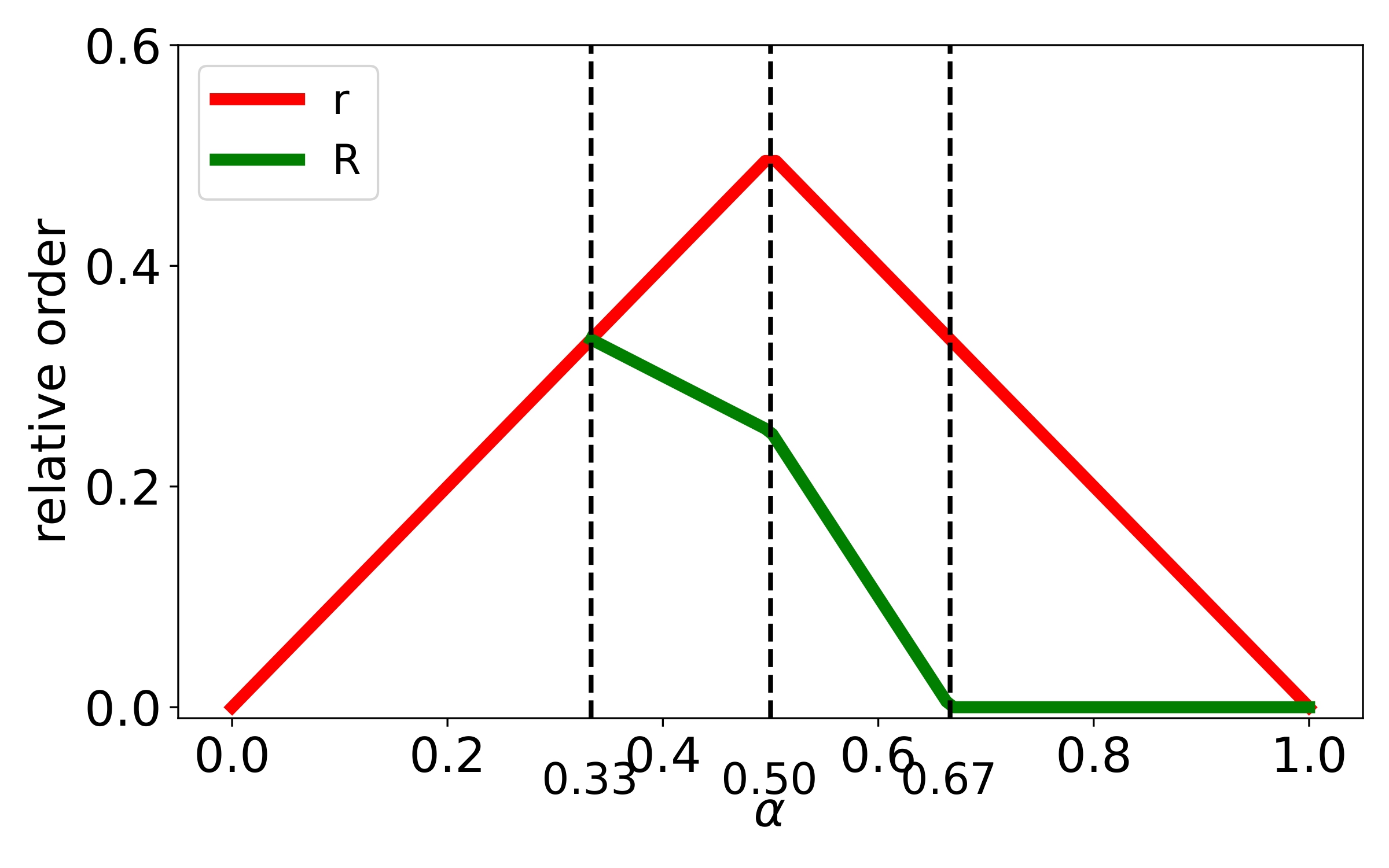}
  \caption{Relative order of convergence between LDP-ASGD and LDP-SGD estimators.}
  \label{fig:summary}
\end{figure}

We present a summary of our theoretical findings in Table \ref{tab1}. This subsection is dedicated to discussing the optimal choice for the step size decay rate. Notice that for any \(\alpha \in (0,1)\), the LDP-ASGD estimator exhibits a faster convergence rate compared to the LDP-SGD estimator. We define the relative order of convergence between the LDP-ASGD and LDP-SGD estimators for the parameter as \( r = \lim_{n \to \infty} \left\{ -\log_n \left( \bar{\delta}_n / \delta_n \right) \right\} \), which quantifies the asymptotic improvement in the convergence rate achieved by the averaged estimator. For example, if \( \delta_n = \mathcal{O}(n^{-\alpha}) \) and \( \bar{\delta}_n = \mathcal{O}(n^{-2\alpha}) \), then \( r = \lim_{n \to \infty} \left\{ -\log_n \left( \bar{\delta}_n / \delta_n \right) \right\} = \lim_{n \to \infty} \left\{ -\log_n (C n^{-\alpha}(1 + o(1))) \right\} = \alpha \). Similarly, the relative order of convergence for the loss is defined as \( R = \lim_{n \to \infty} \left\{ -\log_n \left( \bar{\Delta}_n / \Delta_n \right) \right\} \). The relationship between the base exponent \( \alpha \) and the relative orders \( r \) and \( R \) is illustrated in Figure~\ref{fig:summary}.

\begin{table}[h]
\fontsize{11.5}{16.5}\selectfont
\centering
\renewcommand{\arraystretch}{1}
\scalebox{0.9}{\begin{tabular}{|c|c|c|c|c|}
\hline
   $\alpha$            & $(0,1/3]$     &  $[1/3,1/2]$  & $[1/2,2/3]$   & $[2/3,1)$     \\ \hline
$\delta_n$                &   $n^{-\alpha}$     &   $n^{-\alpha}$     &  $n^{-\alpha}$      &     $n^{-\alpha}$        \\ \hline
$\bar{\delta}_n$                &  $n^{-2\alpha}$      &    $n^{-2\alpha}$    &   $n^{-1}$     &      $n^{-1}$            \\ \hline
\end{tabular}}\,\,\scalebox{0.9}{\begin{tabular}{|c|c|c|c|c|}
\hline
   $\alpha$            & $(0,1/3]$     &  $[1/3,1/2]$  & $[1/2,2/3]$   & $[2/3,1)$     \\ \hline
$\Delta_n$                &   -     &   $n^{-(3\alpha-1)/2}$     &  $n^{-\alpha/2}$      &     $n^{-(1-\alpha)}$        \\ \hline
$\bar{\Delta}_n$                &  -      &    $n^{-\alpha}$    &   $n^{-(1-\alpha)}$     &      $n^{-(1-\alpha)}$       \\ \hline
\end{tabular}}
\caption{Summary of convergence rate regrading different step size decay rate $\eta_i=\eta i^{-\alpha}$. The $\delta_n$ and $\bar{\delta}_n$ represent the expected distance between the
proposed estimators and the global optimum for LDP-SGD and LDP-ASGD estimators, respectively, while $\Delta_n$ and $\bar{\Delta}_n$ represent the difference between the loss incurred by the proposed estimators and the optimal loss for LDP-SGD and LDP-ASGD estimators, respectively. }
\label{tab1}
\end{table}

Observe that a higher relative order indicates a more significant advantage for the LDP-ASGD estimators in terms of achieving faster convergence when compared to the LDP-SGD estimators. As illustrated in Figure \ref{fig:summary}, To minimize the parameter distance, setting \(\alpha = 1/2\) is found to be optimal for achieving the highest relative order of \(1/2\),  indicating that the ratio \(\bar{\delta}_n/\delta_n\) scales as \(\mathcal{O}(1/\sqrt{n})\). This selection is consistent with the asymptotic distribution findings presented in Corollary \ref{coro1}. On the other hand, to minimize the loss distance, the best strategy involves selecting the decay rate $\alpha$ to be as low as feasible to guarantee convergence, with the optimal value being \(\alpha = 1/3\). Note that the optimal relative order \( r \) and \( R \) are achieved at different values of \( \alpha \). This discrepancy arises from the fact that the convergence analyses for the parameter estimation error and the expected excess risk are based on different assumptions, and thus the two quantities are not directly comparable. In particular, under non-strongly convex settings, the minimizer \( \theta^* \) may not be unique, making it infeasible to analyze the convergence rate of the parameter error in a meaningful way.

 \section{Simulation studies}\label{sec:experiment}
 \renewcommand{\theequation}{5.\arabic{equation}}
\setcounter{equation}{0}

In this section, we illustrate and validate our theoretical results through two numerical simulations. In particular, we show the behavior of LDP-SGD and LDP-ASGD estimators in both strongly and non-strongly convex cases. To evaluate the results, we define the parameter distance of an estimator \( \hat{\theta} \) to the optimum \( \theta^* \) as \( \mathbb{E}(\|\hat{\theta} - \theta^*\|^2) \). Similarly, we define the loss distance to the optimum as \( \mathbb{E}(\|F(\hat{\theta}) - F(\theta^*)\|^2) \).


\subsection{Comparison between different private mechanisms}

As mentioned in Remark \ref{rmk1}, Laplace and Gaussian noises exhibit distinct behaviors in DP-SGD algorithms, with the divergence largely attributed to their varying sensitivity to dimensionality \( d \). Specifically, given the same boundedness condition on the gradient of the loss function, i.e., Condition \ref{Con1}, Laplace noise exhibits a higher order dependency \( \mathcal{O}(d^3) \) on dimensionality than Gaussian noise  \( \mathcal{O}(d) \).  It is crucial to recognize that Condition \ref{Con1} constrains the \( \ell_2 \)-norm, not the \( \ell_1 \)-norm, of the gradient. This distinction is important because for any \( \theta \in \mathcal{R}^d \), the inequality \( \|\theta\|_2 \leq \|\theta\|_1 \leq \sqrt{d}\|\theta\|_2 \) holds, indicating that constraining the \( \ell_2 \)-norm constitutes a less stringent assumption.

In this subsection, we aim to compare the estimation performance of Laplace and GDP mechanisms while achieving approximately the same level of privacy protection. Consider a linear regression model given by \(y_i = x_i^{\top} \beta + \epsilon_i\), where the disturbances  \(\epsilon_1, \ldots, \epsilon_n\) are independently and identically drawn from \(N\left(0, \sigma^2\right)\). Here, the covariates are denoted as \(x_i=(1, z_i)^{\top}\), with \(z_i,\,i=1,\ldots,n\) is an i.i.d. sample from \(N\left(0, \sigma_{z}^2\mathbb{I}_{d}\right)\). The loss function we have chosen is expressed as:
\[
\mathcal{L}_n(\beta, \sigma) = \frac{1}{n} \sum_{i=1}^n \left(\sigma \rho_c\left(\frac{y_i-x_i^{\top} \beta}{\sigma}\right) + \frac{1}{2} \kappa_c \sigma\right) w\left(x_i\right),
\]
where \(\rho_c\) denotes the Huber loss function with a tuning parameter \(c\), \(\kappa_c\) is a constant chosen to ensure the consistency of \(\hat{\sigma}\), and \(w(x) = \min (1, {2}/{\|x\|_2^2})\) serves to de-weight outlying covariates. By this construction, the global sensitivities of \(\nabla_\theta \mathcal{L}_n(\theta)\) is \(\sqrt{8 c^2+ c^4/4}\). We generate data utilizing a sample size of \( n = 300000 \) based on the linear model, with the true parameters assigned as \( \beta = \boldsymbol{1}_{d+1} \), \( \sigma = 2 \), \( \sigma_z = 1 \). In addition, we adopt \( c = 1.345 \)  for the loss function.

\begin{figure}[ht]
\centering
  \includegraphics[width=0.9\textwidth]{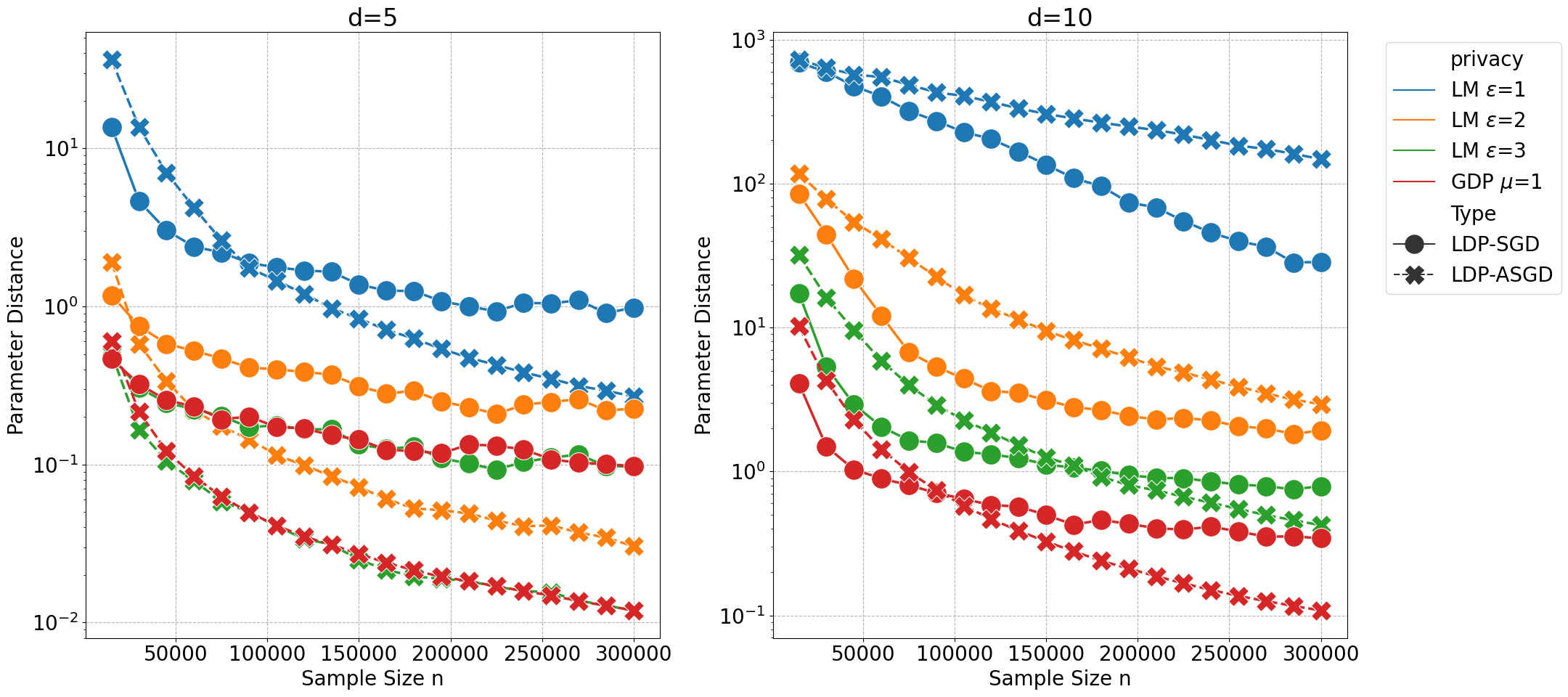}
  \caption{Non-asymptotic distance comparison between Laplace Mechanism (LM) with $\epsilon=1,2,3$, and GDP Mechanism for $\mu=1$ and $d=5,10$.}
  \label{fig:compare}
\end{figure}

Our goal is to assess the effectiveness of LDP-SGD and LDP-ASGD estimators under Laplace and GDP mechanisms. For the GDP mechanism, we set \(\mu=1\). As illustrated in Example \ref{eg3}, this setting naturally results in achieving \((1,0.1269)\)-LDP, \((2,0.0209)\)-LDP, and \((3,0.0015)\)-LDP. We then explore Laplace mechanisms that meet \(1\)-LDP, \(2\)-LDP, and \(3\)-LDP criteria and compare their respective \(\delta_n\) and \(\bar{\delta}_n\) values with those of \(1\)-GDP. Additionally, we fix other parameters, such as the initial step size at \(\eta=0.2\), and a decay rate of \( \alpha = 1/2 \), across two dimensions, \(d=5\) and \(d=10\). For each case, we conduct 200 replications and plot the evolution of the distance between the estimators and the optimal value throughout the training. The outcomes are depicted in Figure \ref{fig:compare}.

As depicted in Figure \ref{fig:compare}, for \(d=5\), the estimator conforming to \(1\)-GDP and satisfying \(3\)-LDP exhibit roughly equivalent parameter distances while providing similar levels of privacy protection. However, at \(d=10\), despite the \(1\)-GDP and \(3\)-LDP estimators maintaining comparable privacy safeguards, the GDP mechanism significantly outperforms the Laplace mechanism with a reduced parameter distance. These findings corroborate our theoretical analysis, highlighting the GDP mechanism's superior robustness to dimensionality relative to the Laplace mechanism.

\subsection{Linear regression}
In thisand the subsequent subsections, we will concentrate on the GDP mechanism due to its advantageous handling of dimensions. We begin by revisiting the linear model and data generation process described in the previous subsection. Our main goal is to perform a parameter sensitivity analysis for the LDP-SGD and LDP-ASGD estimators in the context of the GDP mechanism. To achieve this, we carry out experiments in various settings: parameter dimensions \( d = 5, 10, 20 \);  initial step size $\eta=0.2$ with decay rates \( \alpha = 1/3, 1/2, 2/3, 1 \); privacy budgets \( \mu = 0.5, 1, 2, 3 \); and a non-private SGD approach (without the addition of noise). For each configuration, we perform 200 replications and plot the trajectories of the distance between the estimators and the true parameter values, as well as the loss distance, throughout the training process. The results with \( d = 5 \) are illustrated in Figure \ref{fig:1}, and detailed outcomes for other settings are provided in the Appendix.


\begin{figure}[ht]
  \centering
  \includegraphics[width=0.9\textwidth]{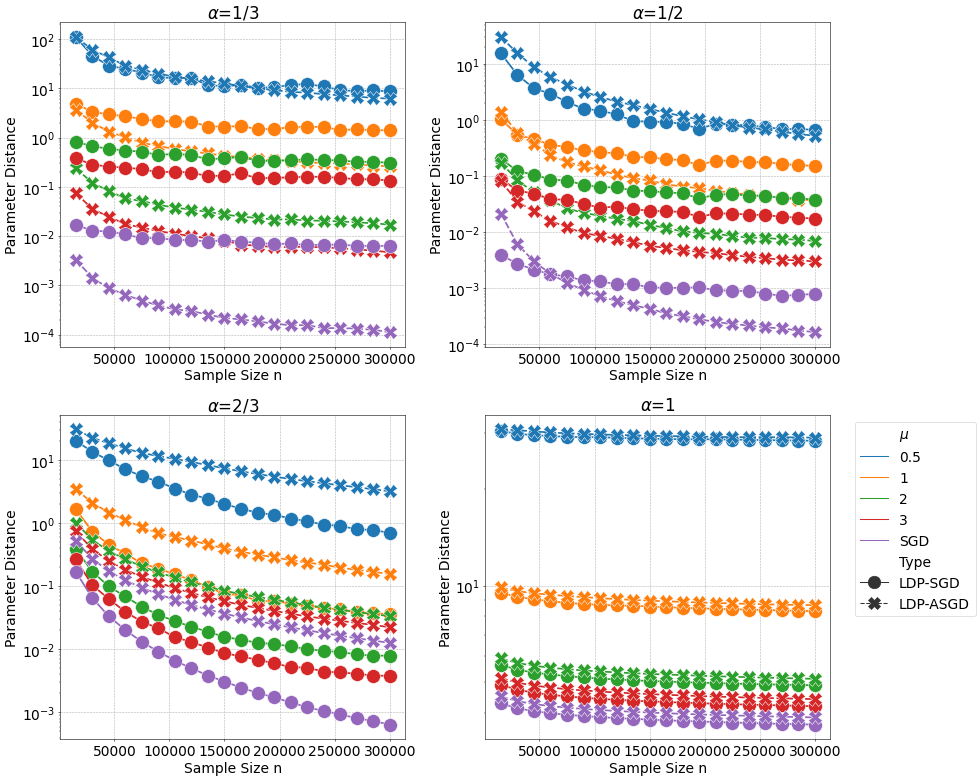}
  \caption{Trajectories of the distance between DP-SGD estimators and the optimal for linear regression with $d=5$.}
  \label{fig:1}
\end{figure}


Figure \ref{fig:1} clearly demonstrates that the empirical results align well with our theoretical predictions. For \( \alpha \in (0, 1/2] \), the LDP-ASGD estimator shows a more rapid convergence than the LDP-SGD estimator, leading to better performance across various privacy settings. In the case of \( \alpha \in [2/3, 1) \), although the LDP-ASGD estimator still converges faster than the LDP-SGD estimator, the constant terms in the upper bound for the LDP-ASGD estimator may exceed those of the LDP-SGD estimator, yielding a greater distance. The constant terms in the upper bound can be found in the supplementary material. It is noteworthy that at \( \alpha = 1 \), neither estimator achieves convergence. Additionally, decreasing the privacy budget \( \mu \) correlates with an increase in distance, a consequence of adding more noise in each iteration.



\subsection{Logistic regression}
In the context of the logistic regression, data are generated based on the model $y_i \sim \operatorname{Bernoulli}( \{1+\exp (-x_i^{\top} \beta )\}^{-1} )$. This procedure employs the same value of \( \beta \) and adopts the same design for generating the covariates \( x_i \) as was applied in the linear regression simulation. We utilize a weighted variant of the standard cross-entropy loss:
\begin{align*}
\mathcal{L}_n(\beta) =& \frac{1}{n} \sum_{i=1}^n ( -y_i \log ( \frac{1}{1+\exp (- x_i^{\top} \beta )} )- ( 1-y_i ) \log ( \frac{\exp (- x_i^{\top} \beta )}{1+\exp ( -x_i^{\top} \beta )} ) ) w( x_i ),
\end{align*}
where the weight function is denoted as  \( w( x ) = \min ( 1, {2}/{\| x \|_2^2})\).

We generate data with a sample size of \( n = 300000 \) drawn from the logistic model, with the true parameters set as \( \beta = \boldsymbol{1}_{d+1} \) and \( \sigma_z = 1 \). 
It is important to note that, in the context of logistic regression, the assumption of strong convexity is no longer applicable. Consequently, our primary focus shifts toward the disparity between the loss incurred by the estimators and the optimal loss. To further assess the robustness of Theorems \ref{the1} and \ref{the2} in scenarios void of strong convexity, we also present results pertaining to parameter distance. The results for \( d = 5 \) are depicted in 
Figure \ref{fig:3}. Detailed results for alternative configurations are provided in the Appendix.


\begin{figure}[ht]
  \centering
  \includegraphics[width=0.9\textwidth]{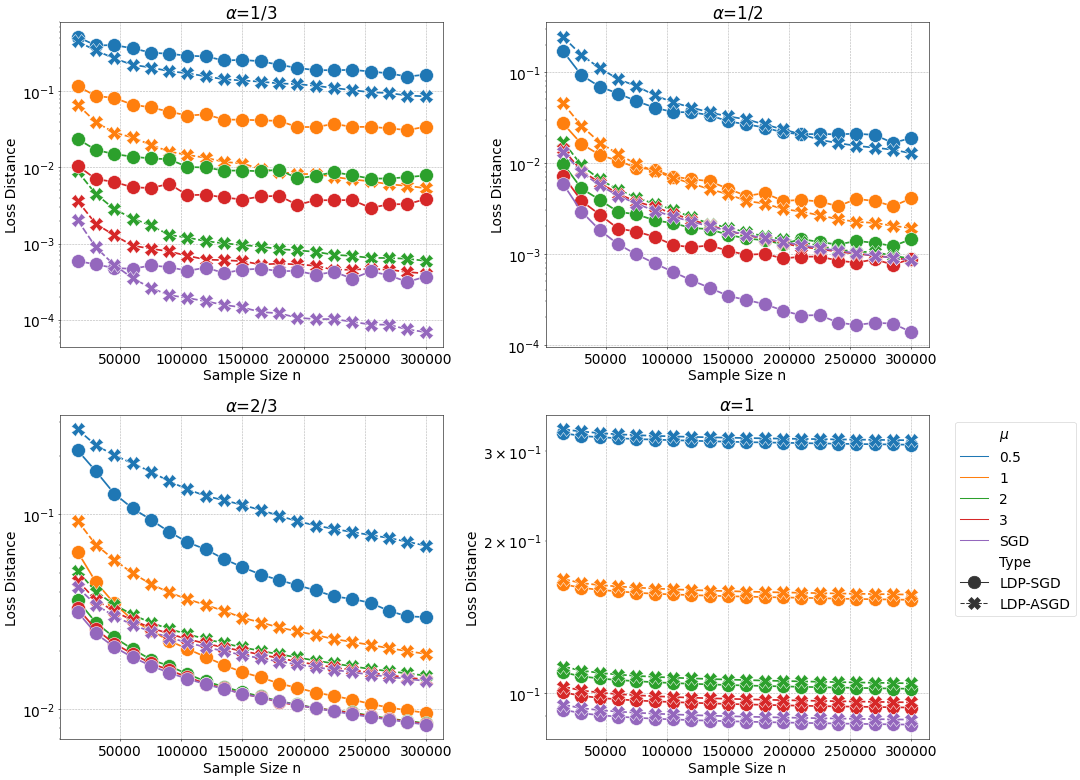}
  \caption{Trajectories of the distance between the loss incurred by the estimators and the optimal loss for logistic regression with $d=5$.}
  \label{fig:3}
\end{figure}

The insights gleaned from Figure \ref{fig:3} align with the theoretical predictions presented in Theorems \ref{the3} and \ref{the4}. Specifically, for any \( \alpha \in [1/3,2/3] \), the LDP-ASGD estimator achieves faster convergence than the LDP-SGD estimator. However, when \( \alpha > 2/3 \), despite having the same convergence rate, the constant term associated with the LDP-ASGD estimator may become more pronounced. This could potentially lead to a larger loss distance for the averaged estimators.


\section{US insurance data}
\label{sec:real}

{{
In this section, we demonstrate the proposed LDP-SGD and LDP-ASGD algorithms using the synthetic data based on the US insurance dataset, which includes health insurance premium charge records in the United States. The  dataset captures various factors influencing medical costs, including age, gender, body mass index (BMI), number of children, smoking status, region, medical history, exercise frequency, occupation, and type of insurance plan. It was generated by randomly sampling 1,000,000 records to provide a comprehensive representation of the insured population. The dataset is publicly available at \textsf{https://www.kaggle.com/datasets/sridharstreaks/insurance-data-for-machine-learning}.

For the analysis, extensive data preprocessing was performed to ensure proper handling of both numerical and categorical variables. Numerical variables, such as age, BMI, and the number of children, were standardized to have a mean of zero and a standard deviation of one, ensuring consistency across features for modeling. Categorical variables were encoded with meaningful numeric values to reflect their underlying semantics. Health-related variables, such as \textit{medical\_history} and \textit{family\_medical\_history}, were encoded based on condition severity: \textit{None} (0), \textit{High blood pressure} (1), \textit{Diabetes} (2), and \textit{Heart disease} (3). The \textit{exercise\_frequency} variable was encoded as \textit{Never} (0), \textit{Occasionally} (1), \textit{Rarely} (2), and \textit{Frequently} (3) to reflect varying activity levels. For \textit{occupation}, hierarchical encoding was applied, ranging from \textit{Student} (0) to \textit{White collar} (3), capturing the occupational spectrum. Similarly, the \textit{coverage\_level} variable was encoded as \textit{Basic} (0), \textit{Standard} (1), and \textit{Premium} (2) to represent the varying levels of insurance coverage. Binary variables, such as \textit{smoker} and \textit{gender}, were encoded as 0 and 1 to represent \textit{no}/\textit{yes} or \textit{female}/\textit{male}, respectively.

In the following, we consider two settings:
\begin{enumerate}
    \item Linear regression: we regard ``charges'' as the numerical response and use other features to predict it. The loss function is chosen as the Huber loss with Mallows weight defined in Section 5.1.
    \item Multinomial Logistic regression: we divide ``charges'' into three classes based on its quantiles at $\tau=0.33$ and $\tau=0.66$, then regard it as the categorical response, and use other features to classify it. We use the cross-entropy loss with Mallows weight through the training process as defined in Section 5.2.
\end{enumerate}

In our experimental framework, we begin by dividing the total dataset of one million records into two parts: 800,000 for training and 200,000 for testing. During the iterative training process, the training data is sequentially fed into the model, with the performance of both LDP-SGD and LDP-ASGD estimators on the test dataset monitored at every 10000th iteration. We set the initial step size $\eta=0.5$ and explore a range of $\alpha$ values as mentioned in Section 4.2. The parameters are initialized by sampling from the uniform distribution Unif$(0,1)$. To ensure robustness and account for variability in our data processing, the experiment is repeated 50 times with random permutations of the training data to vary the input order for the online algorithm. The results are aggregated and analyzed, with the average performance for each configuration presented in the subsequent figures. For linear regression, we present the loss verses sample size under different configurations, while for logistic regression, we display both loss and accuracy $P(y=\hat{y})$, where $y$ is the groud truth label and $\hat{y}$ denotes the predicted label, verse sample size for different settings.

The linear regression results are shown in Figure \ref{fig:11}. The figure demonstrates that the proposed LDP-SGD and LDP-ASGD algorithms perform well, with a low MSE comparable to the non-private SGD version when $\mu \geq 1$. The estimators become stable when $n = 10,000$ for $\mu = 2$ and $\mu = 3$, as indicated by the flattened MSE loss. However, a stronger privacy guarantee (lower $\mu$) leads to slower convergence, as the loss decreases more gradually. Additionally, the loss for the LDP-ASGD estimator is consistently higher than that of the LDP-SGD estimators when $\alpha\geq1/2$. This occurs because the averaged estimator requires more steps to forget the initial values while the sample size of the dataset is relatively small. The phenomenon supports the same convergence properties as discussed in the Theorem \ref{the1} and \ref{the2}, and demonstrated in the simulation studies. When $\alpha = 1/2$, the LDP-ASGD estimator attains the lowest loss across all privacy budget settings compared to other step size decay rates $\alpha$, indicating that the optimal convergence rate is achieved at $\alpha = 1/2$. When $\alpha=1$, both LDP-SGD and LDP-ASGD estimators fail to converge.

\begin{figure}[ht]
  \centering
  \includegraphics[width=0.9\textwidth]{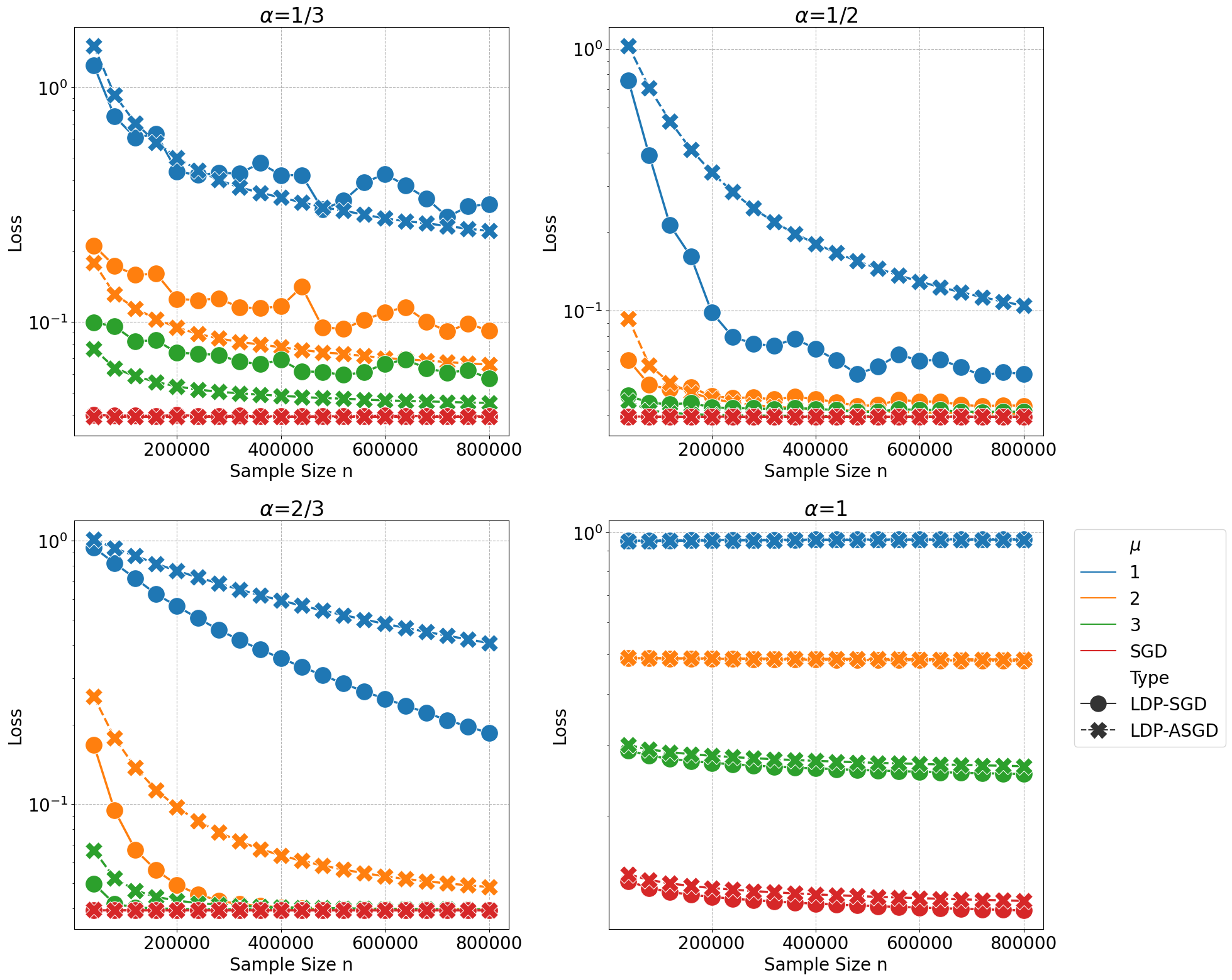}
  \caption{Trajectories of the loss on test dataset for US insurance dataset using linear regression with Huber loss.}
  \label{fig:11}
\end{figure}

The results for logistic regression are shown in Figures \ref{fig:12}-\ref{fig:13}. These figures demonstrate that the proposed LDP-SGD and LDP-ASGD algorithms perform well, with both cross-entropy loss and prediction accuracy approaching the levels of the non-private SGD version. As with previous findings, the loss for the LDP-ASGD estimator is consistently higher than for LDP-SGD, due to the limited sample size. The algorithms exhibit the same convergence properties discussed in the Theorem \ref{the3} and \ref{the4}. When $\alpha = 1/3$, the LDP-SGD estimator achieves the lowest loss across all privacy budget settings, indicating the optimal convergence rate. However, when $\alpha = 1$, both LDP-SGD and LDP-ASGD fail to converge.

\begin{figure}[ht]
  \centering
  \includegraphics[width=0.9\textwidth]{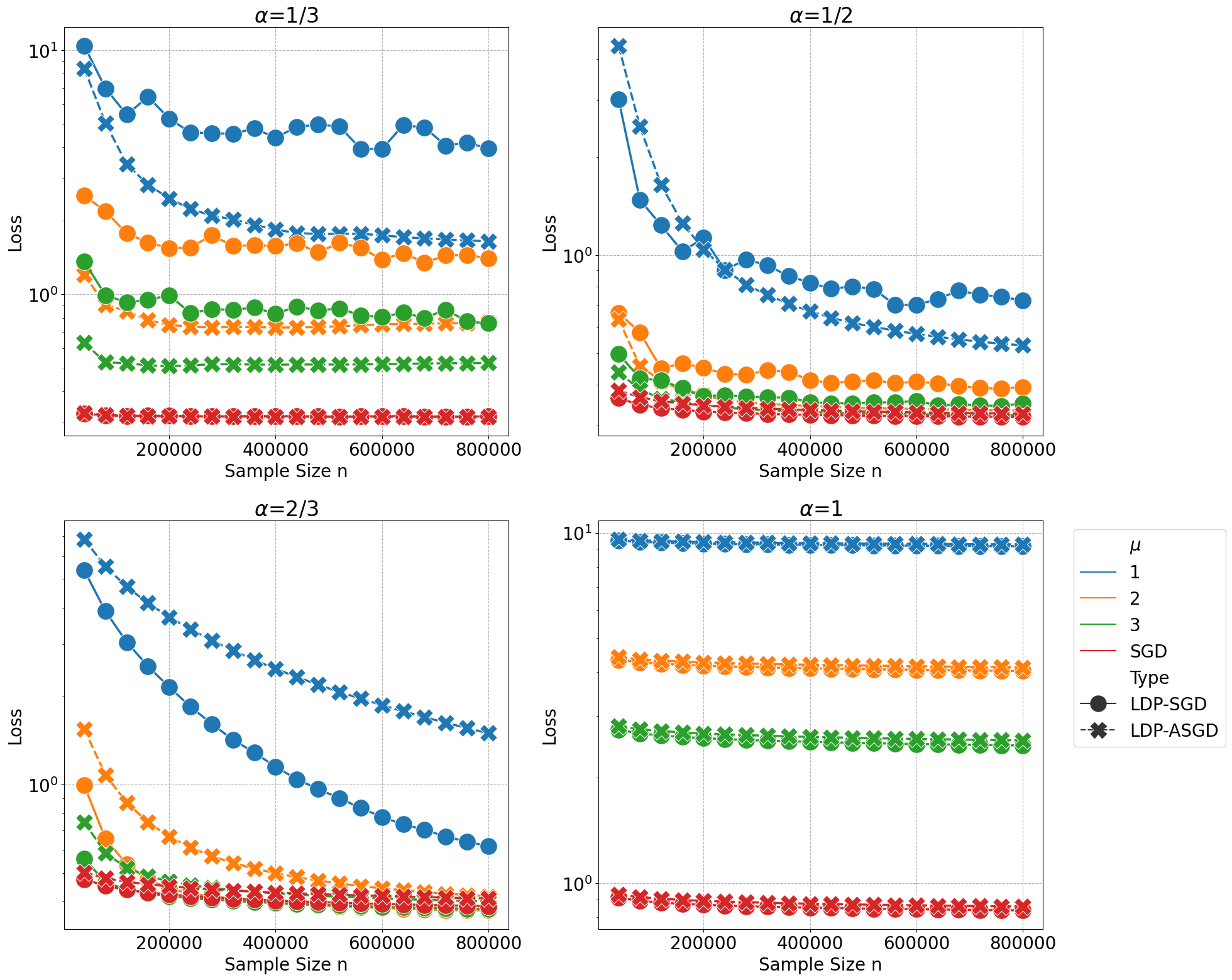}
  \caption{Trajectories of the cross-entropy loss on test dataset for US insurance dataset using logistic regression.}
  \label{fig:12}
\end{figure}

\begin{figure}[ht]
  \centering
  \includegraphics[width=0.9\textwidth]{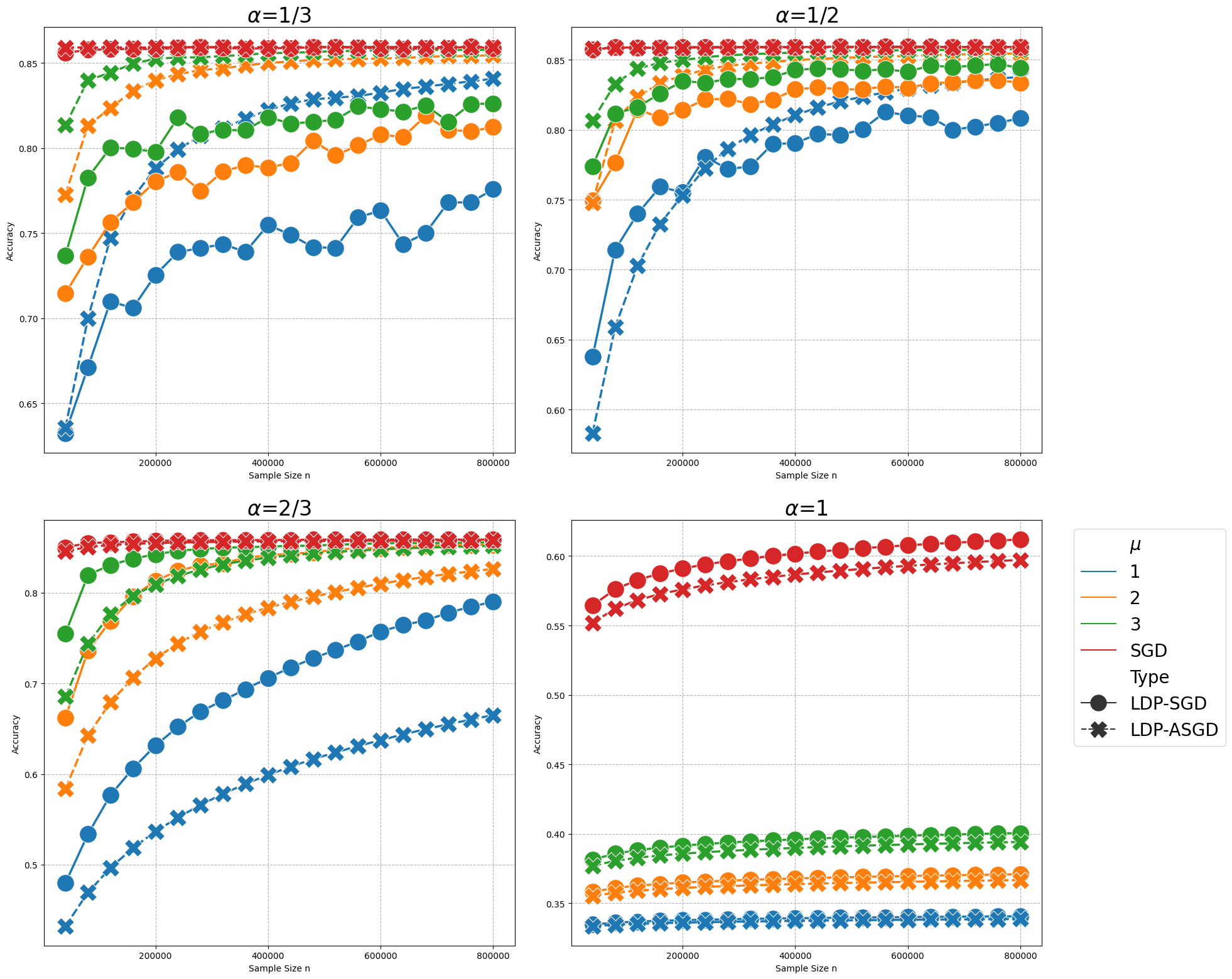}
  \caption{Trajectories of the classification accuracy on test dataset for US insurance dataset using logistic regression.}
  \label{fig:13}
\end{figure}

}}

\section{Conclusion}
\label{sec:con}
In this paper, we have provided a non-asymptotic analysis of the convergence of the general LDP-SGD algorithm, as well as its averaged version under the LDP model, which allows different individual users to have different privacy budgets. Importantly, we have analyzed the theoretical properties of the proposed estimators, both with and without strong convexity assumptions. Our theory shows how the convergence rates of the considered estimators are affected by various hyperparameters, including step size, parameter dimensions, and privacy budgets. Our results provide useful guidelines on how the choice of the hyper-parameters may impact the results. Numerical studies yield affirmation of the asymptotic results.

\newpage

\bibliography{ref}
\bibliographystyle{chicago}

\end{document}